\documentclass[preprint]{elsarticle}
\usepackage[dvipsnames,usenames]{xcolor}
\usepackage{graphicx}
\usepackage{dcolumn}
\usepackage{bm}
\usepackage{mathrsfs,natbib}
\usepackage{graphicx,epsf,amssymb,amsbsy,amsfonts,amssymb,amsmath}
\usepackage{slashed}
\usepackage{comment}
\usepackage{subcaption}
\usepackage{ulem} 
\usepackage{endnotes}
\usepackage[margin=3.0cm]{geometry}
\newcommand{\eq}[1]{Eq.~(\ref{#1})}

\usepackage[hidelinks]{hyperref}
\hypersetup{
    colorlinks=true,       
    linkcolor=black,          
    citecolor=blue,        
    filecolor=magenta,      
    urlcolor=blue           
}

\newcommand{\beq}{\begin{eqnarray}}
\newcommand{\eeq}{\end{eqnarray}}

\begin{document}
\begin{frontmatter}

\title{Finite-Temperature Quarkyonic Matter \\ with an Excluded Volume Model}

\author{Srimoyee Sen}
\ead{srimoyee08@gmail.com}
\address{Department of Physics and Astronomy, Iowa State University, Ames, IA 50011-3160}
\author{Neill C. Warrington}
\ead{ncwarrin@uw.edu}
\address{Institute for Nuclear Theory, University of Washington, Seattle, Washington 98195-1550}

\begin{abstract}
We introduce a theoretical framework to analyze the thermal properties of a recently proposed ``excluded volume" model of quarkyonic matter. This entails proposing  finite temperature distribution functions and entropy functionals constrained by internal theoretical consistency. We use the formalism developed here to analyze the effect of temperature on the quark onset as a function of baryon density. We find that the baryon density at which the quarkyonic phase emerges decreases only slightly with temperature, an effect produced by large $N_c$. We furthermore find that a finite temperature increases the density of quarks produced. The framework presented here can be used to compute the equation of state of hot and dense quarkyonic matter which is of relevance to neutron star mergers.
\end{abstract}

\end{frontmatter}


\section{Introduction} 
Quarkyonic matter is a proposed phase of intermediate density QCD at large $N_c$ where confinement persists \cite{McLerran:2007qj}. In this phase, baryons form a degenerate Fermi gas where the total baryon number is distributed among both quark and nucleon degrees of freedom. One postulated feature of quarkyonic matter \cite{McLerran:2007qj, Kojo:2009ha} that sets it apart from other mixed quark-nucleon phases is the structure of the momentum space distribution function. At zero temperature, due to confining forces at the Fermi surface, the momentum space distribution consists of an inner sphere of quarks surrounded by an outer shell of nucleons. Compared with models exhibiting first order phase transitions \cite{PhysRevLett.122.061101,Oechslin:2004yj}, this shell structure produces a stiff equation of state at moderate densities, while retaining a soft equation of state at lower densities, making quarkyonic matter attractive for neutron star phenomenology \cite{McLerran:2018hbz,Fukushima:2015bda,Jeong:2019lhv}. 

It was recently shown \cite{McLerran:2018hbz} that a pronounced peak in the speed of sound of nuclear matter, expected from neutron star observations \cite{Tews_2018}, is reproduced by a model of quarkyonic matter. In this model, the momentum space shell structure is imposed through particular choices of quark and nucleon distribution functions. Such a momentum space shell structure cannot so far be established directly from QCD, so a natural question that follows is whether one can generate such a distribution function dynamically from another model. Indeed such a model was recently found in \cite{Jeong:2019lhv} where quarks and nucleons both appear as quasi-particles and nucleons interact with each other via hard core repulsion. We will refer to this model as the ``excluded volume" model of quarks and nucleons for the rest of the paper\footnote{Some recent papers that describe nuclear interactions with excluded volume model include \cite{Bugaev:2018uum, Blaschke:2020qrs,Bugaev:2020sgz, Vovchenko:2017drx}}. Minimizing the total energy density of this two-component model yields a hadronic distribution function at low total baryon density and a quarkyonic distribution function at high baryon density \cite{McLerran:2018hbz}. 

In this paper we build a framework that captures thermal properties of the excluded volume model of \cite{Jeong:2019lhv}. This involves proposing theoretically consistent thermal distribution functions and entropy functionals which generate expected behaviors of quarkyonic matter at finite temperature. By theoretically consistent we mean that the finite temperature thermodynamics produced by the framework here has the correct limiting behavior, e.g. converging exactly to the thermodynamics of Fermi sphere of all nucleon and all quark Fermi configurations. Note that the excluded volume model proposed in \cite{Jeong:2019lhv}, and extended here, deliberately excludes detailed nuclear interactions beyond the hard-core repulsion of the excluded volume. This is because such details are not central to the framework being explored in this paper, and only serve as a distraction. Operating under these assumptions, our purpose is to analyze the effect of temperature. It is important to remember that the excluded volume model considered in this paper employs massive quarks, and, in its current state, cannot capture the detailed patterns of chiral symmetry restoration and di-quark condensation predicted in Ginzburg-Landau free-energy calculations of high density QCD in the chiral limit \cite{Hatsuda_2006}.

 A primary motivation of our analysis is probing the existence of quark matter through binary neutron star mergers \cite{Chatziioannou:2019yko}. Recently it has become evident that information about the phase of matter present may be contained in the post-merger signal \cite{Bauswein:2018bma,PhysRevLett.122.061101}, and some questions which may be addressed are: whether neutron stars have deconfined quark matter in their cores \cite{Chatziioannou:2019yko}, and whether there is a phase transition between hadronic and quark degrees of freedom \cite{PhysRevD.100.066027}. Reaching temperatures on the order of $50$ MeV \cite{Perego:2019aa}, however, neutron star mergers present new theoretical challenges. Most ab-initio methods in nuclear physics are designed for zero temperature systems \cite{SCHMIDT199999,PhysRevC.97.044318,RevModPhys.87.1067}, and analyzing mergers with lattice QCD is currently not practical due to the sign problem \cite{deForcrand:2010ys}. In absence of an exact method, few model calculations of finite temperature dense matter have been performed \cite{PhysRevD.100.066027}. The primary goal of this paper, therefore, is to contribute to these few studies by extending the excluded volume model of quarkyonic matter model to finite temperatures. In order to demonstrate the utility of the framework proposed here, we compute within this model, the effect of a finite temperature on the quark-onset density i.e. the baryon density at which a purely nucleonic distribution function gives way to a quarkyonic distribution function. We find that the effect of raising the temperature, though small, is to favor quark production.

This paper is organized as follows. In Sec. \ref{zero-temperature-theory} we will define the excluded volume model at zero-temperature. In Sec. \ref{finite-temperature-theory} we extend the model to finite temperature and in Sec. \ref{results} we present our findings. Finally, we conclude in Sec. \ref{conclusion}.

\section{Excluded Volume Model at $T=0$} 
\label{zero-temperature-theory}
 In this section we review the zero temperature $(T=0)$ excluded volume model proposed in \cite{Jeong:2019lhv}. In particular, we explain how the distribution function of baryons evolves from purely nucleonic to quarkyonic as the baryon density is increased. For simplicity, throughout this paper we will consider isospin symmetric matter.
 
\begin{figure}[t!]
\begin{center} 
	\includegraphics[scale=0.5]{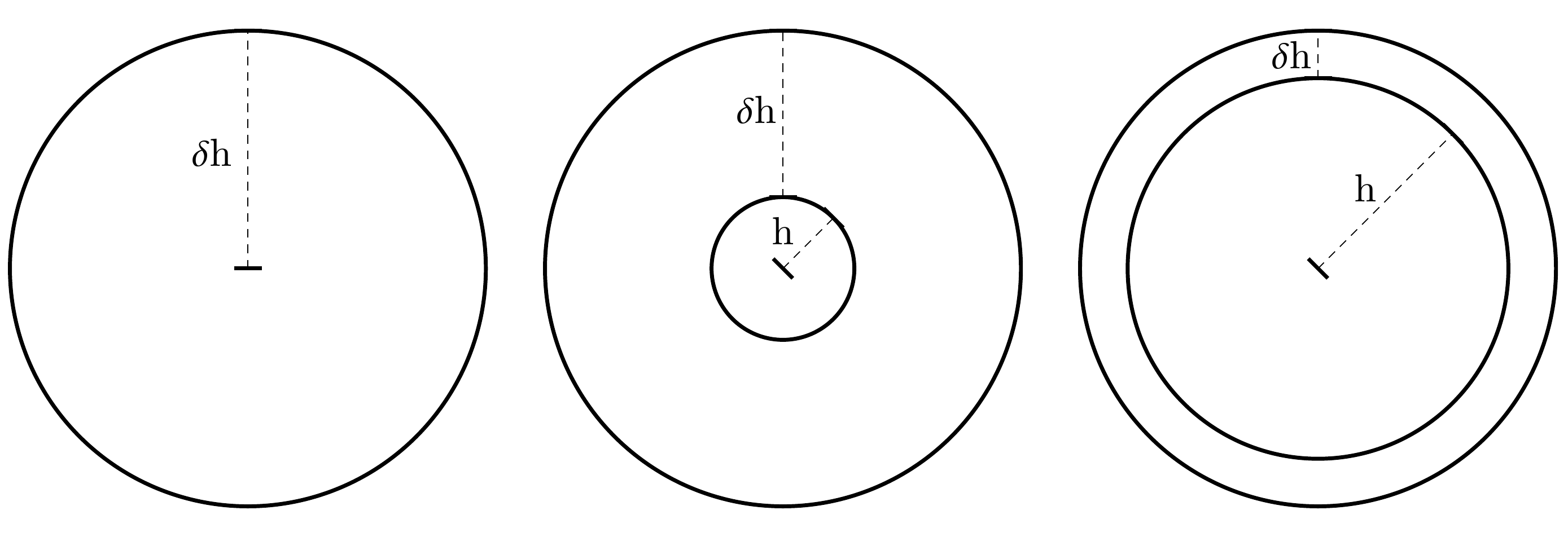} 
\end{center}
\caption{Here we depict two-dimensional projections of the momentum space shell structure of quarkyonic matter at a variety of total baryon number densities, increasing from left to right. At low density, the system is composed entirely of nucleons, resulting in a momentum space distribution which is all ``shell", with radius $\delta h$. Increasing the density, an inner Fermi surface of quarks with radius $h$ is produced. The inner Fermi sphere grows with density, leading eventually to a thin shell of nucleons in momentum space.}
\label{fig:shell-structure}
\end{figure}

 The central idea of quarkyonic matter is that confinement persists even at high baryon density \cite{McLerran:2007qj}. Due to confining forces, quarks near the edge of the Fermi sea bind together to produce nucleons, while quarks deep within the Fermi sea, being Pauli blocked, do not. These considerations lead to a momentum space ``shell structure", depicted in Fig. \ref{fig:shell-structure}, where high density matter is supposed to be a Fermi sphere of free quarks surrounded by a Fermi shell of nucleons. 
 
While the existence of quarkyonic matter cannot currently be deduced from first principle calculations in QCD, the excluded volume model of \cite{Jeong:2019lhv} generates the shell structure dynamically at zero temperature. The excluded volume model includes both nucleons and quarks as degrees of freedom, with quarks treated as free particles and nucleons interacting via hard-core repulsion. The hard-core repulsion is modeled by assigning each nucleon a volume $v_0$, so that $N$ nucleons in volume of $V$ can only access a volume of $V - N v_0$. This implementation of repulsive forces is perhaps crude, however such forces do exist, as can be deduced from partial wave analyses \cite{PhysRevC.48.792} and chiral effective field theory \cite{Epelbaum:2008ga}. In addition to producing the shell structure, the excluded volume theory predicts sensible low and high density limits: at low baryon density one obtains a dilute gas of nucleons, and at high density a dense gas of quarks. 

To implement the model, note that both quarks and nucleons contribute to the total baryon number density:
\beq
n_{tot}=n_N+n_Q~.
\eeq
Here $n_{tot}$ is the total baryon number density, and $n_N$ and $n_Q$ are the baryon number density in nucleons and quarks, respectively \footnote{Note that the baryon number of a single quark is $1/N_c$ and that there are $N_c$ species of quarks in our theory. These facts lead to non-trivial cancellations in the thermodynamic functionals.}. Related to $n_N$ and $n_Q$ are the energy density in nucleons and quarks, $\epsilon_N$ and $\epsilon_Q$. According to the laws of thermodynamics \cite{callen1960thermodynamics}, at zero temperature the equilibrium configuration is obtained by minimizing the total energy density 
\beq
\epsilon_{\text{tot}} = \epsilon_N + \epsilon_Q
\eeq
at fixed total baryon density. The equilibrium configuration of the system is fully specified through this minimization procedure when $\epsilon_N(n_N)$ and $\epsilon_Q(n_Q)$ are supplied. In the excluded volume model, these dependencies are supplied through intermediary quantities, namely the distribution functions of quarks and nucleons, which we denote as $f_Q(k)$ and $f_N(k)$, and take to be:
\begin{align}
f_Q(k) & = \theta(h/N_c - \epsilon_Q(k)) \nonumber \\
f_N(k) & = \theta(h + \delta h - \epsilon_N(k)) \theta(\epsilon_N(k)-h)~.
\label{eq:dist-func}
\end{align}
These definitions implement a Fermi sphere of quarks surrounded by a shell of nucleons. The baryon density in quarks $n_Q$ and the baryon density in nucleons $n_N$ are defined as the following functionals of $f_Q$ and $f_N$ \cite{Jeong:2019lhv}:
\begin{align}
n_Q & = N_f (2s+1)\int{\frac{d^3 k}{(2 \pi)^3} f_Q(k)} \nonumber \\ 
\frac{n_N}{1-n_N/n_0} & = N_f (2s+1)\int{\frac{d^3 k}{(2 \pi)^3} f_N(k)}~, 
\label{eq:dens-func}
\end{align}
as are the energy densities:
\begin{align}
\epsilon_Q & = N_c N_f (2s+1)\int{\frac{d^3 k}{(2 \pi)^3} f_Q(k)~\epsilon_Q(k)} \nonumber \\
\frac{\epsilon_N}{1-n_N/n_0} & = N_f (2s+1)\int{\frac{d^3 k}{(2 \pi)^3} f_N(k)\epsilon_N(k)}~.
\label{eq:energy-func}
\end{align}
Here $\epsilon_Q(k)$ and $\epsilon_N(k)$ are free quark and nucleon dispersion relations, $n_0 \equiv v_0^{-1}$ is the hard-core density, $N_f=2$ and $s=1/2$ is the spin of a quark or nucleon. We neglect throughout this paper anti-particle contributions to all thermodynamic quantities, as they are suppressed by powers of the vanishingly small quantity $e^{-2M/T}$, when $T\simeq 50$ MeV. The distribution functions \eq{eq:dist-func}, as well as the thermodynamic functionals \eq{eq:dens-func}, \eq{eq:energy-func} define the $T=0$ excluded volume model.

\begin{figure}[t!]
\begin{center}
	\includegraphics[scale=0.5]{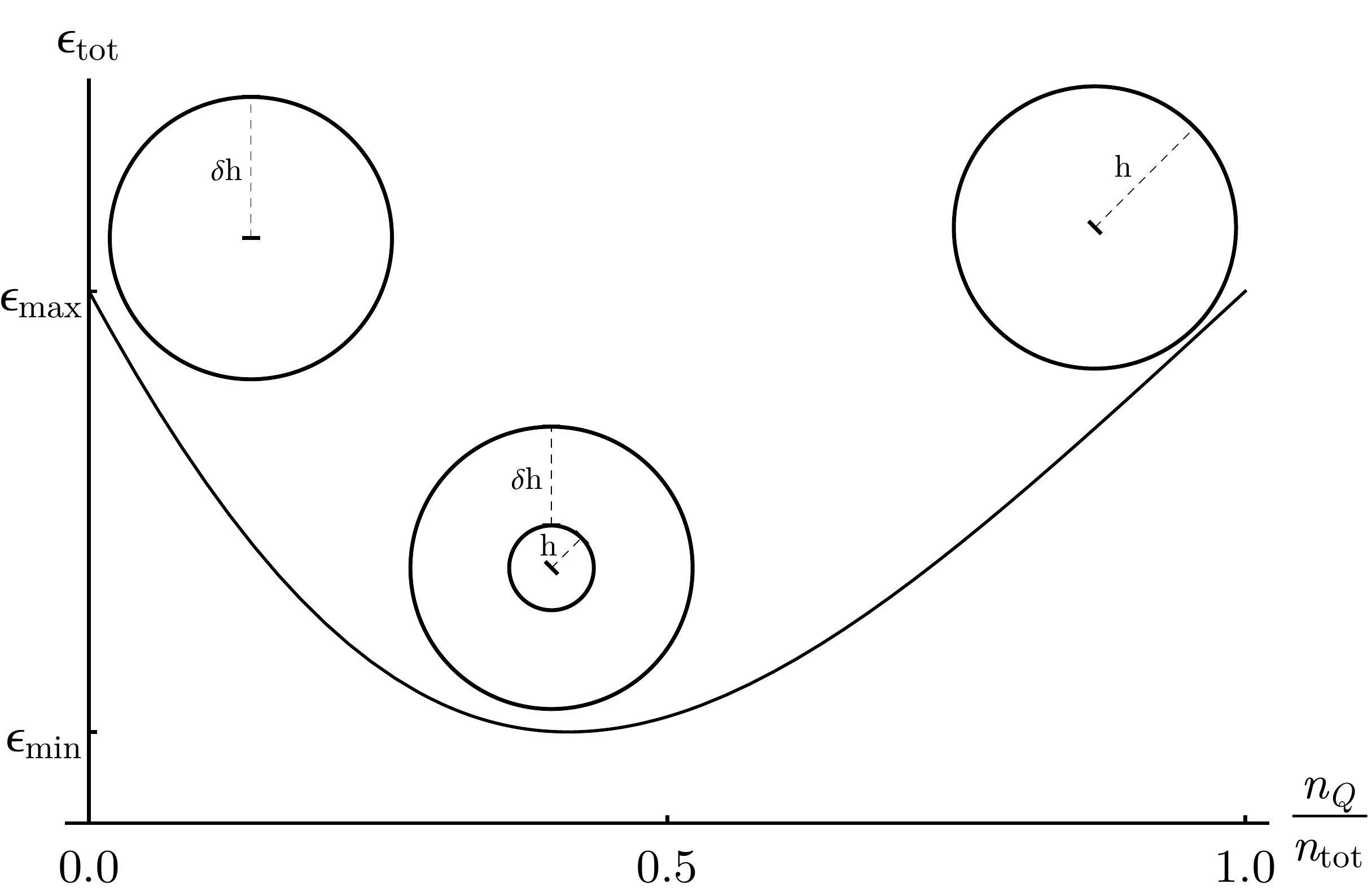}
\end{center}
\caption{Here we schematically depict the energy minimization procedure. The horizontal axis is the fraction of baryon density in quarks, and the vertical axes is the total energy density for a given quark fraction. The width of the shell varies as the quark fraction is varied. The system's equilibrium state is that which minimizes the total energy density.}
\label{fig:energy-min-schematic}
\end{figure}

For intuition of how the model might capture the properties of high density matter, note that baryon number can be stored either in quark degrees of freedom or nucleon degrees of freedom. No particular configuration is enforced, and which degrees of freedom are actually present in equilibrium at a given density is determined by the energy cost. This is depicted schematically in Fig. \ref{fig:energy-min-schematic}. 

We now demonstrate that the excluded volume theory generates quarkyonic matter. To proceed, we must specify the dispersion relations of the quarks and nucleons. Throughout this paper we employ non-relativistic dispersion relations:
\begin{align}
\epsilon_Q(k) & = m + \frac{k^2}{2 m} \nonumber \\
\epsilon_N(k) & = M + \frac{k^2}{2 M}~,
\end{align}
where $M$ is the nucleon mass and $m$ is the quark mass with $m\equiv M/N_c$. These dispersion relations reasonably well-approximate the  relativistic dispersion relations of free quarks and nucleons at the densities found in neutron stars and neutron star mergers, where the density spans the range $0\lesssim n/n_{sat}\lesssim 5$ \cite{Perego:2019aa,Radice:2016rys} \footnote{We have used both relativistic and non-relativistic dispersion relations in our numerical calculations and find nearly identical results under the thermodynamic conditions analyzed here.}. Throughout we define $n_{sat}\equiv 0.16~ \text{fm}^{-3}$ to be nuclear saturation density. With these specifications, 
\begin{align}
n_Q & = N_f (2s+1)\int_{0}^{\sqrt{2m(h/N_c -m)}}{\frac{d^3 k}{(2 \pi)^3}} \nonumber \\ 
\frac{n_N}{1-n_N/n_0} & = N_f (2s+1)\int_{\sqrt{2M(h -M)}}^{\sqrt{2M(h+\delta h -M)}}{\frac{d^3 k}{(2 \pi)^3}}~, 
\end{align}
and 
\begin{align}
\epsilon_Q & = N_f (2s+1)\int_{0}^{\sqrt{2m(h/N_c -m)}}{\frac{d^3 k}{(2 \pi)^3}\Big(m + \frac{k^2}{2m}\Big)} \nonumber \\ 
\frac{\epsilon_N}{1-n_N/n_0} & = N_f (2s+1)\int_{\sqrt{2M(h -M)}}^{\sqrt{2M(h+\delta h -M)}}{\frac{d^3 k}{(2 \pi)^3}\Big(M + \frac{k^2}{2M}\Big)}~.
\end{align}
The total energy density is then given by
\beq
\epsilon_{tot} - M n_{tot}= N_f (2s+1)\Bigg[\int_{0}^{\sqrt{2m(\mu/N_c -m)}}{\frac{d^3 k}{(2 \pi)^3} \frac{k^2}{2m}} + (1-n_N/n_0)\int_{\sqrt{2M(\mu -M)}}^{\sqrt{2M(\mu+\delta \mu -M)}}{\frac{d^3 k}{(2 \pi)^3} \frac{k^2}{2M}}\Bigg]~. \nonumber \\
\eeq
The $M n_{tot}$ term is produced by the combined contributions of the rest mass energy from quarks and nucleons, and does not affect a minimization at fixed $n_{tot}$. We therefore drop the rest mass contribution in all subsequent calculations. 
\begin{figure}[t!]
\begin{center} 
	\includegraphics[scale=0.65]{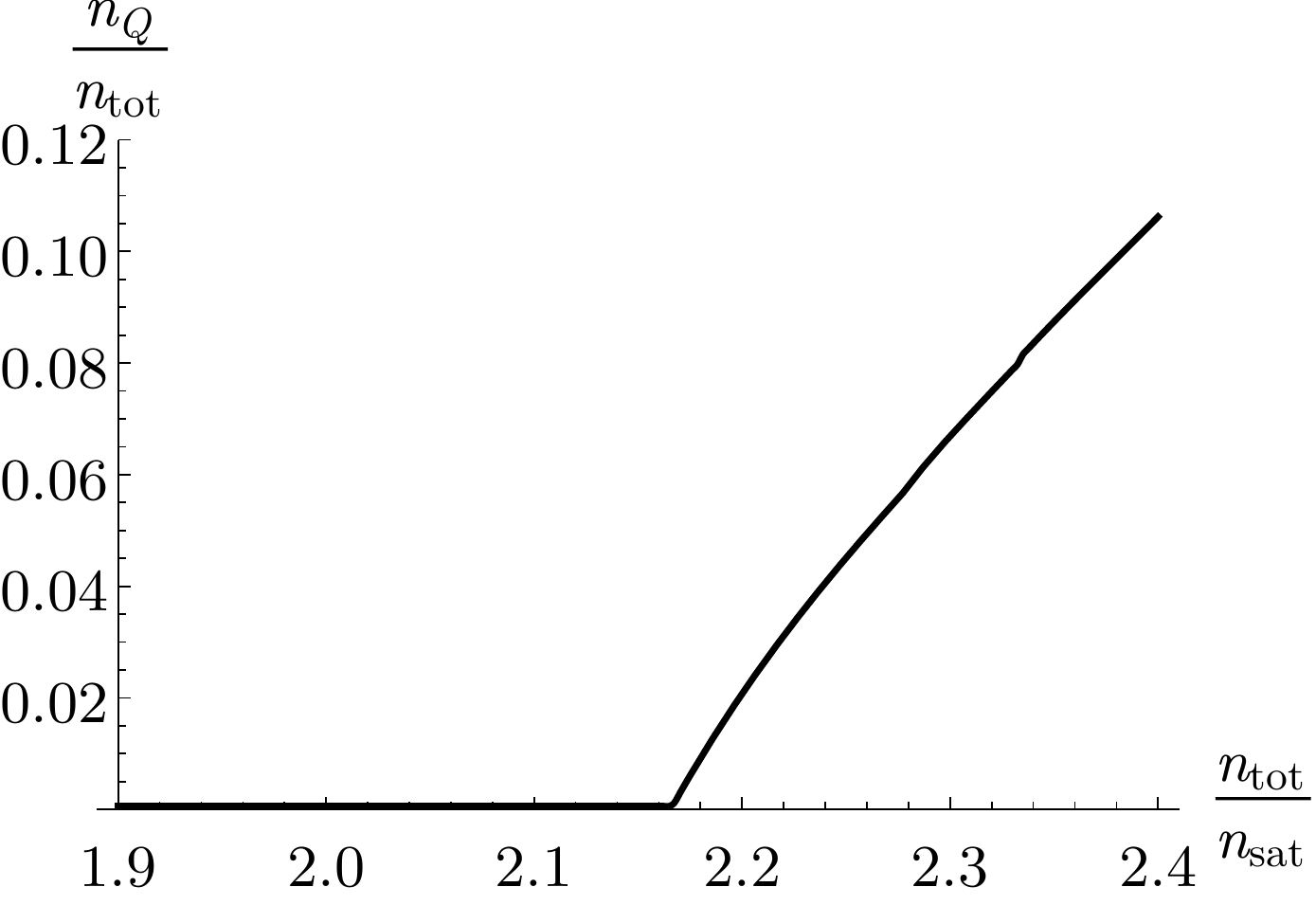}
\end{center}
\caption{Here we plot the equilibrium quark fraction $n_Q/n_{tot}$ as a function of total baryon density. At low density, only nucleons are found in the system while at high densities quarks begin to be produced. }
\label{fig:zero-temp-quark-fraction}
\end{figure}
It is straightforward to numerically compute the integrals above, then minimize the total energy at fixed $n_{tot}$. Following this procedure with the hard core density to be $n_0 = 2.5 n_{sat}$ results in Fig. \ref{fig:zero-temp-quark-fraction}. Here it is seen that the excluded volume model dynamically produces quarkyonic matter: at low density the system is composed of only nucleons, but around the hard-core density quarks begin to be dynamically produced. 

Due to the presence of the two Fermi surfaces, however, seeing this behavior analytically is complicated. To understand the quark onset behavior, it is useful to look at two limiting cases: all nucleons or all quarks. In either of these cases, the two Fermi surfaces collapse to one, simplifying expressions. For a given total baryon density $n_{tot}$, the energy of an all quark configuration is
\beq
\epsilon_Q = \frac{1}{5\pi^2 M}N_c^2\left(\frac{3\pi^2 n_{tot}}{2}\right)^{5/3}.
\label{eq:allq}
\eeq
while the energy of an all nucleon configuration is
\beq
\epsilon_N=\frac{1}{5\pi^2 M}\left(\frac{3\pi^2n_{tot}}{2}\right)^{5/3}\frac{1}{\left(1-n_{tot}/n_0\right)^{2/3}}.
\label{eq:alln}
\eeq
These are qualitatively different behaviors, and account for the general features of Fig. \ref{fig:zero-temp-quark-fraction}. At low density, when $n_{tot}<<n_0$, the energy required to produce quarks is roughly $N_c^2$ times the energy needed to produce baryons. This is why at low densities the equilibrium configuration consists of all nucleons. However, the energy required to produce nucleons increases with density due to the repulsive interactions, and when 
\beq
\left(1-n_{tot}/n_0\right)^{2/3}\sim N_c^2
\eeq
nucleons become more difficult to produce than quarks. At this point, the density of nucleons saturate, while quarks are produced in ever greater amount.

\section{Excluded Volume quarkyonic matter at finite temperature}
\label{finite-temperature-theory}

\begin{figure}[t!]
\begin{center}
	\includegraphics[scale=0.7]{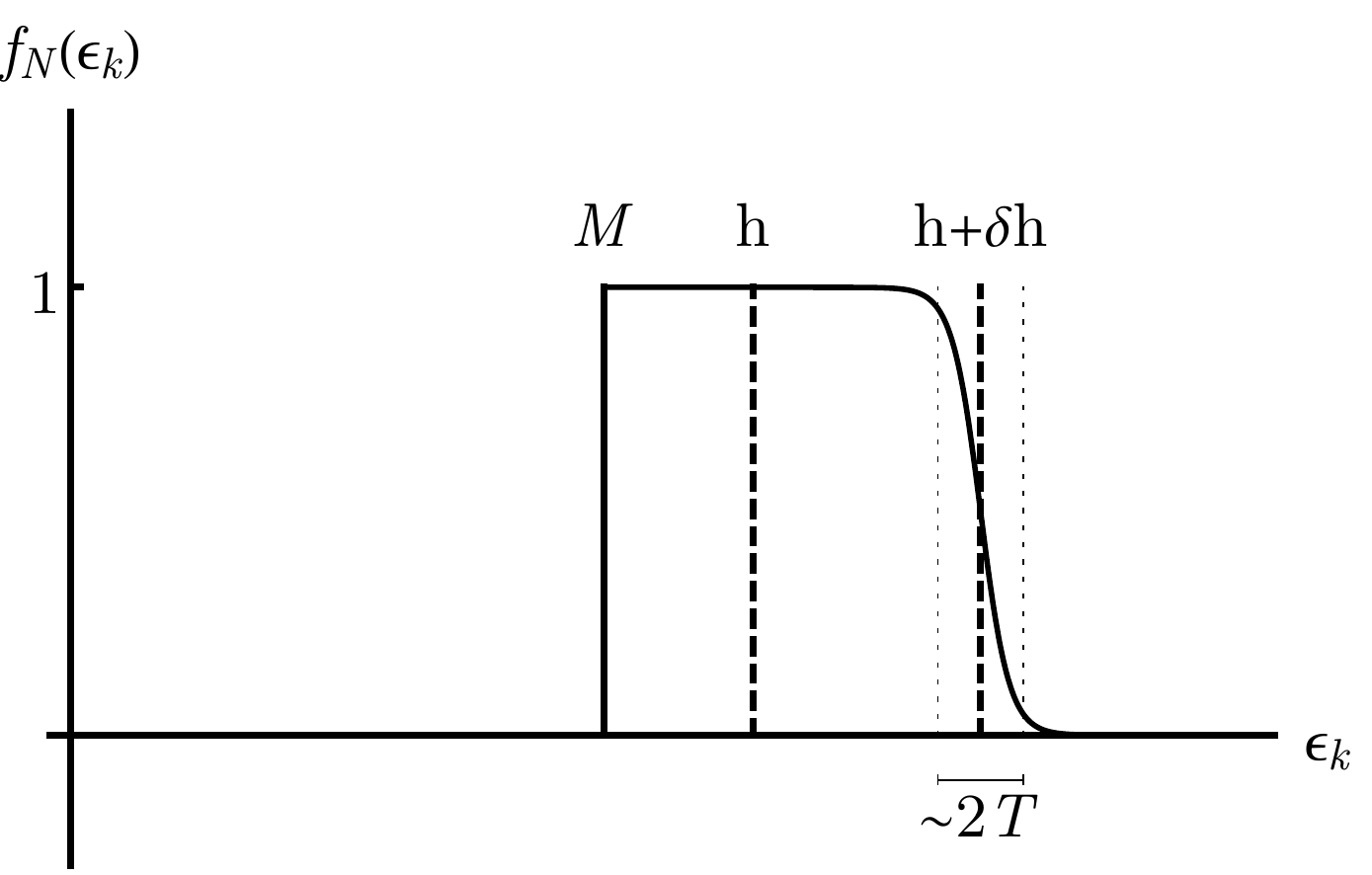}
\end{center}
\caption{Here we plot the distribution functions defined in \eq{eq:dist-func-nonzeroT} as a function of energy. Support for the distribution function begins at $M$ and persists until roughly $h + \delta h$ where the Fermi surface is smeared out over a width of order $T$. The energy band $M\leq \epsilon \leq h$ encodes the quark distribution, while the band $h\leq \epsilon\leq h + \delta h$ encodes the nucleon distribution.}
\label{fig:finite-temp-dist-func}
\end{figure}

We now introduce a framework for describing quarkyonic matter at finite temperature. In order to capture thermal effects we have to first remember that the quarkyonic distribution function is designed to minimize the effects of thermal fluctuations on the quarks. The motivation behind this is that the lowest energy degrees of freedom for a degenerate Fermi gas lie at the Fermi surface and such degrees of freedom for the quakryonic phase are nucleons by construction. Any thermal distribution function for the quarkyonic matter has to be consistent with this feature. This requirement, combined with a smooth thermal smearing of the nucleon and the quark degrees of freedom, leads us to propose the following distribution functions for quarks and nucleons $f_Q$ and $f_N$ respectively:
\begin{align}
f_Q(k) & = g(\epsilon_Q(k), \frac{h+\delta h}{N_c}) \theta(h /N_c - \epsilon_Q(k))  \nonumber \\
f_N(k) & = g(\epsilon_N(k), h+\delta h) \theta(\epsilon_N(k)-h) 
\label{eq:dist-func-nonzeroT}
\end{align}
where $g$ is the Fermi-Dirac distribution
\beq
g(\epsilon,\mu) = \frac{1}{e^{\beta(\epsilon-\mu)}+1}~,
\eeq
and $\beta=T^{-1}$. These distribution functions are plotted in Fig. \ref{fig:finite-temp-dist-func} for a visual aid. The density and energy density functionals are still taken to be \eq{eq:dens-func} and \eq{eq:energy-func}, but are now computed with the distribution functions \eq{eq:dist-func-nonzeroT}. 

This choice of $f_Q$ and $f_N$, with a common envelope $g$, implements both the shell structure (through the $\theta$ functions) and the observation that there is just one underlying distribution function in quarkyonic matter. Since there is just one outermost Fermi surface, heating the system by a small amount should result in a deformation only at this edge (at $h + \delta h$). One feature which is evident from Fig. \ref{fig:finite-temp-dist-func} is that nucleons are affected by temperature before quarks. In the following section we will see this feature produces particular dynamics

Entropy is a new ingredient entering the finite temperature theory. Just as internal energy functionals proposed in \cite{Jeong:2019lhv} are needed at $T=0$, entropy functionals are needed at $T\neq0$. This is because at $T\neq0$ and fixed baryon density, it is the Helmholtz free energy, rather than the internal energy, which is minimized in equilibrium. Since the Helmholtz free energy density $F$ is related to the energy density $\epsilon$ by $F=\epsilon - T s$ ($s$ is the entropy density and $T$ is the temperature), a finite temperature description will require entropy functionals. We will denote the quark and nucleon contributions to the Helmholtz free energy density as $F_Q$ and $F_N$, and the total Helmholtz free energy density as $F_{tot}$. We take the entropy functionals for quarks ($s_Q$) and nucleons ($s_N$) to be:
 \begin{align}
 s_Q  & =  N_f N_c(2s+1) \int{\frac{d^3 k}{(2 \pi)^3}\theta( h/N_c-\epsilon_Q(k))\Big[\beta \xi_Q(k) f_Q(k) + \text{ln}(1+e^{-\beta \xi_Q(k)})\Big]} \nonumber \\
  \frac{s_N}{1-n_N/n_0}  & =  N_f (2s+1) \int{\frac{d^3 k}{(2 \pi)^3}\theta(\epsilon_N(k) - h)\Big[\beta \xi_N(k) f_N(k) + \text{ln}(1+e^{-\beta \xi_N(k)})\Big]} 
  \label{eq:entropy-density}
 \end{align}
where $\xi_{Q}(k)=\epsilon_Q(k) - (h+\delta h)/N_c$ and $\xi_{N}(k)=\epsilon_N(k) - (h +\delta h)$. 
To understand this choice of $s_Q$ and $s_N$, recall that the excluded volume model consists of $N_c$ species of free quarks within the Fermi sea, on top of which is a Fermi shell of nucleons whose interactions are completely encoded by an excluded volume (i.e the nucleons are free within the excluded volume). Our aim is to implement precisely these observations. Given that the entropy density for a single species for free fermion is 
 \beq
 s = -\frac{1}{V}\frac{\partial F}{\partial T} = \int{\frac{d^3k}{(2 \pi)^3} \Big[\beta (\epsilon(k) - \mu ) f(k)+ \text{ln}(1+e^{-\beta (\epsilon(k) - \mu )})\Big]}~,
 \eeq
where $f(k)$ is the Fermi-Dirac distribution, we see that the proposed $s_Q$ and $s_N$ describe the entropy of free fermions with the shell structure imposed. The remaining factors in $s_Q$ and $s_N$ account for powers of $N_c$ and the fact that nucleons interact in an excluded volume.

\vspace{-3.5pt}
\section{Results}
\label{results}
In this section we use the framework proposed in the previous section to compute thermodynamic quantities. To achieve this we have to find, given a fixed total baryon density, the distribution functions $f_Q$ and $f_N$ which minimize the Helmholtz free energy density. Once these functions are known, all thermodynamic quantities can be computed. To accomplish this goal, first a total density $n_{tot}$ is chosen, and is partitioned into $n_Q$ and $n_N$ such that $n_Q+n_N=n_{tot}$. Next, the density functionals \eq{eq:dens-func} are inverted to produce $h$ and $\delta h$. These parameters are then used to compute the energy and entropy densities (\eq{eq:energy-func} and \eq{eq:entropy-density}). This procedure is repeated over all partitions. Minimizing the free energy density, the equilibrium values of $h$ and $\delta h$ (and hence the distribution functions) are obtained.

In order to demonstrate the utility of our model, we focus on the effect of temperature on quark production. In particular, we compute the onset baryon density of quarks as a function of temperature as well as the equilibrium quark fraction. We first employ a low-temperature expansion \cite{Sommerfeld:1928aa}, where the expansion parameter is $\frac{T}{n_{\text{tot}}^{2/3}/M}$, to understand thermal effects analytically. This expansion is useful to build an intuition for the numerical results that follow. The low temperature expansion, as will be explained, suggests that heating the system aids in quark production. Relative to the zero temperature case, heating the system reduces the onset baryon density by a small amount and raises the fraction of quarks present in equilibrium. We then elaborate upon these analytic arguments with exact numerical calculations.

Before we delve into the analytic calculation in the low temperature limit, it is important to emphasize that a generic quarkyonic configuration with a finite Fermi momentum for the quarks and finite nucleon shell width is cumbersome to analyze analytically even in the low temperature expansion. Therefore, to develop intuition we will analyze two configurations that are composed of only nucleons and of only quarks. We call these configurations ``all-nucleon" and ``all-quark" configurations. A similar exercise was conducted at zero temperature in \cite{McLerran:2018hbz} in order to build intuition for the zero temperature analysis. The numerical results will of course take into account all possible configurations.

To understand the thermal properties of all-quark and all-nucleon configurations in a low temperature expansion, we first write their density, energy density and the entropy functionals  in terms of their Fermi energies. For an all quark configuration, the nucleon shell width is $\delta h=0$ since there are no nucleons in the system, and the Fermi sphere is produced entirely by the quark Fermi energy $h$. Similarly, for an all nucleon configuration, the quark Fermi energy $h=0$ and the Fermi sphere is produced entirely by the nucleon Fermi shell $\delta h$. To implement the low temperature expansion we will expand in the parameters $T/h$ and $T/\delta h$. Note that, the expansion parameter $T/\delta h$ and $T/ h$ are both equivalent to an expansion in $\frac{T}{n_{tot}^{2/3}/M}$. This is because in an all nucleon or an all quark configuration, the Fermi energies are proportional to $n_{tot}^{2/3}/M$. Retaining terms only up to the first nontrivial order in the expansion parameter, the density, and the energy density and the entropy density for an all-nucleon configuration can be written as
\begin{align}
\frac{n_N}{1-n_N/n_0} & = (2s+1)N_f \frac{(2 M \delta h)^{3/2}}{6 \pi^2}\Bigg[1+\frac{\pi^2}{8}\left(\frac{T}{\delta h}\right)^2\Bigg] \nonumber \\
\frac{\epsilon_N}{1-n_N/n_0} & =(2s+1)N_f \delta h\frac{(2 M \delta h)^{3/2}}{10 \pi^2}\Bigg[1+\frac{5 \pi^2}{8}\left(\frac{T}{\delta h}\right)^2\Bigg] \nonumber \\
\frac{s_N}{1-n_N/n_0} & = (2s+1)N_f\frac{(2 M \delta h)^{3/2}}{12} \left(\frac{T}{\delta h}\right) ~.
\label{eq:N}
\end{align}
In order to do the same for quarks in a concise way, we will define the variable $h'\equiv \frac{h - M}{N_c}$. 
We can then write the corresponding quantities for the all-quarks configuration in the low temperature expansion as:
 \begin{align}
n_Q & = (2s+1) N_f \frac{(2 m h')^{3/2}}{6 \pi^2}\Bigg[1+\frac{\pi^2}{8}\big(\frac{T}{h'}\big)^2\Bigg] \nonumber \\
\epsilon_Q & =(2s+1) N_f N_c h'\frac{(2 m h')^{3/2}}{10 \pi^2}\Bigg[1+\frac{5 \pi^2}{8}\big(\frac{T}{h'}\big)^2\Bigg] \nonumber \\
s_Q & = (2s+1) N_f N_c \frac{(2m h')^{3/2}}{12} \big(\frac{T}{h'}\big) ~.
\label{eq:Q}
\end{align}
We use $n_N$ and $n_Q$ to denote the nucleon and quark densities, keeping in mind that we are interested in comparing the free energies of an all quark and an all nucleon configuration; at the end of the calculation we set $n_N = n_Q = n_{tot}$. Our goal is to write the Helmholtz energy density for both configurations as a function of the temperature and the total baryon density. As in the zero temperature case, without loss of generality we drop the rest mass contribution from both the quark and nucleon energy densities. A comparison of the temperature dependence of the energy densities for these two configurations will then reveal whether quarks are favored or disfavored as the temperature is raised. In order to express the energy density as a function of temperature and total baryon density, we invert the equations for $n_Q$ and $n_N$ to write the Fermi energies $h'$ and $\delta h$ for the two configurations as a function of $n_Q$ and $n_N$ respectively. Again, we work up to the leading temperature correction. To lowest order, the Fermi energies read
\begin{align}
\delta h(T)/\delta h(0) & = 1-\frac{{\pi^2}}{12}\big(\frac{T}{\delta h(0)}\big)^2 \nonumber \\
h'(T)/h'(0)& = 1-\frac{{\pi^2}}{12}\big(\frac{T}{h'(0)}\big)^2
\label{eq:h}
\end{align}
where $ \delta h(0)=\frac{1}{2M}\Big(\frac{6 \pi^2 n_N/(1-n_N/n_0)}{N_f (2s+1)}\Big)^{2/3}$ and $h'(0) = \frac{1}{2m}\Big(\frac{6 \pi^2 n_Q}{N_f (2s+1)}\Big)^{2/3}$ are the standard zero temperature Fermi energies of free fermions. Substituting the expressions for $\delta h$ and $h'$ from \eq{eq:h} in the expressions for the entropy and energy densities in \eq{eq:N} and \eq{eq:Q} one finds 
\begin{align}
s_N  & = (1-n_N/n_0)^{2/3} M  n_N^{1/3}T \Big(\frac{(2s+1)\pi N_f}{6}\Big)^{2/3} \nonumber \\
s_Q  & =  m N_c n_Q^{1/3}T \Big(\frac{(2s+1)\pi N_f}{6}\Big)^{2/3}
\end{align}
and 
\begin{align}
\epsilon_N & = \big(\frac{3}{5}\delta h(0) n_N\big)\Big[1+\frac{5}{3 \pi^2}\left(\frac{(2s+1)\pi N_f}{6}\right)^{2/3}\frac{M^2 T^2}{(n_N/(1-n_N/n_0))^{4/3}}\Big] \nonumber \\
\epsilon_Q & = \big(\frac{3}{5}N_c h'(0) n_Q\big)\Big[1+\frac{5}{3 \pi^2}\left(\frac{(2s+1)\pi N_f}{6}\right)^{2/3}\frac{M^2 T^2}{n_Q^{4/3}}\Big] ~.
\end{align}
Combing the energy and entropy we can write down expressions for the Helmholtz free energy for the all-nucleon and all-quark configurations at temperature $T$, denoted as $F_N(T)$ and $F_Q(T)$ respectively. The change in the free energy for an all nucleon configuration due to the introduction of a temperature is
\beq
F_N(T)-F_N(0) = -\frac{1}{2}\left(1-\frac{n_N}{n_0}\right)^{2/3}M n_N^{1/3}T^2\left(\frac{(2s+1)\pi N_f}{6}\right)^{2/3}
\eeq
while the change in free energy for an all quark configuration is 
\beq
F_Q(T)-F_Q(0) = -\frac{1}{2}M n_Q^{1/3}T^2\left(\frac{(2s+1)\pi N_f}{6}\right)^{2/3}~.
\eeq
Now setting $n_N = n_Q = n_{\text{tot}}$, the meaning of these free energy shifts becomes clear. The ratio of (the changes in) the free energy for the two configurations is
\beq
\frac{F_Q(T)-F_Q(0)}{F_N(T)-F_N(0)} = \frac{1}{(1-n_{\text{tot}}/n_0)^{2/3}}~. 
\label{eq:free-energy-shift-estimate}
\eeq
This makes it clear that both configurations decrease in free energy as the temperature is increased, however the all-quark configuration decreases in energy faster than the all-nucleons configuration. Therefore, while the two distribution functions considered correspond to the limiting configurations, it is reasonable to conclude that more quarks will be produced at finite temperature than at zero temperature. This expectation is validated by our numerical calculations. 

\begin{figure}[t!]
\begin{center}
	\centerline{\includegraphics[scale=0.5]{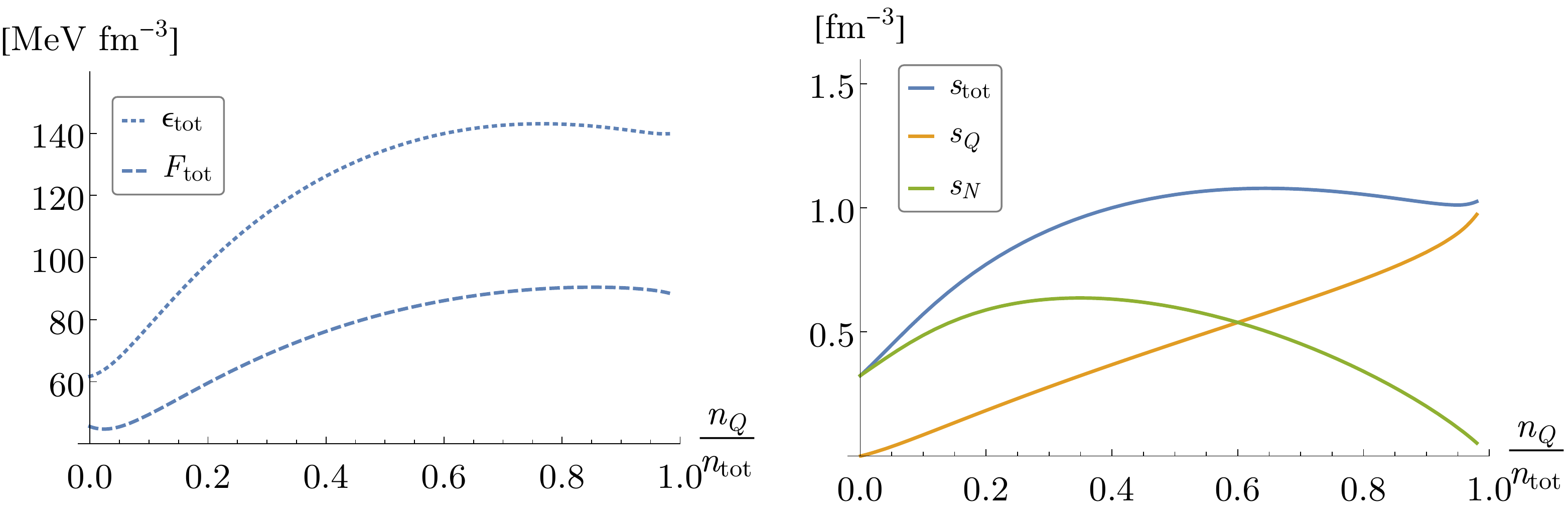}}
\end{center}
\caption{Displayed are typical curves used in the minimization procedure. Here we have taken $n_0=2.5 n_{sat}$, $T=50$ MeV and $n_{tot}=2.2 n_{sat}$. The left panel shows the internal energy and free energy densities as a function of quark fraction, while the right panel shows the individual and total entropies where $s_N$ is the nucleon contribution to the total entropy and $s_Q$ is the quark contribution to the total entropy. The effect of entropy is to shift the minimum of the free energy from zero.}
\label{fig:typical-calculation}
\end{figure}

We now present our numerical results. In all that follows we take $N_c=3$, $M=938\text{ MeV}$, $N_f =2$ and $s=1/2$. To begin, Fig. \ref{fig:typical-calculation} illustrates a typical minimization procedure. Here the hard core density is set to $2.5 n_{sat}$, $T=50$ MeV and $n_{tot}=2.2 n_{sat}$. In the left panel we plot the internal energy and the free energy density as a function of the quark fraction which shows that, while the energy density $\epsilon_{tot}$ has a minimum at zero quark fraction, the total free energy density $F_{tot}$ does not. This shift is due to entropy, shown on the right. As expected from the behavior of free particles, the entropy of quarks increases monotonically with quark fraction. This is not the case with nucleons, where the maximum entropy occurs at a quark fraction of $\sim 30\%$. This non-trivial behavior is due to the competition between increasing density and the excluded volume factor $(1-n_N/n_0)$.

\begin{figure}[t!]
\centerline{\includegraphics[scale=0.5]{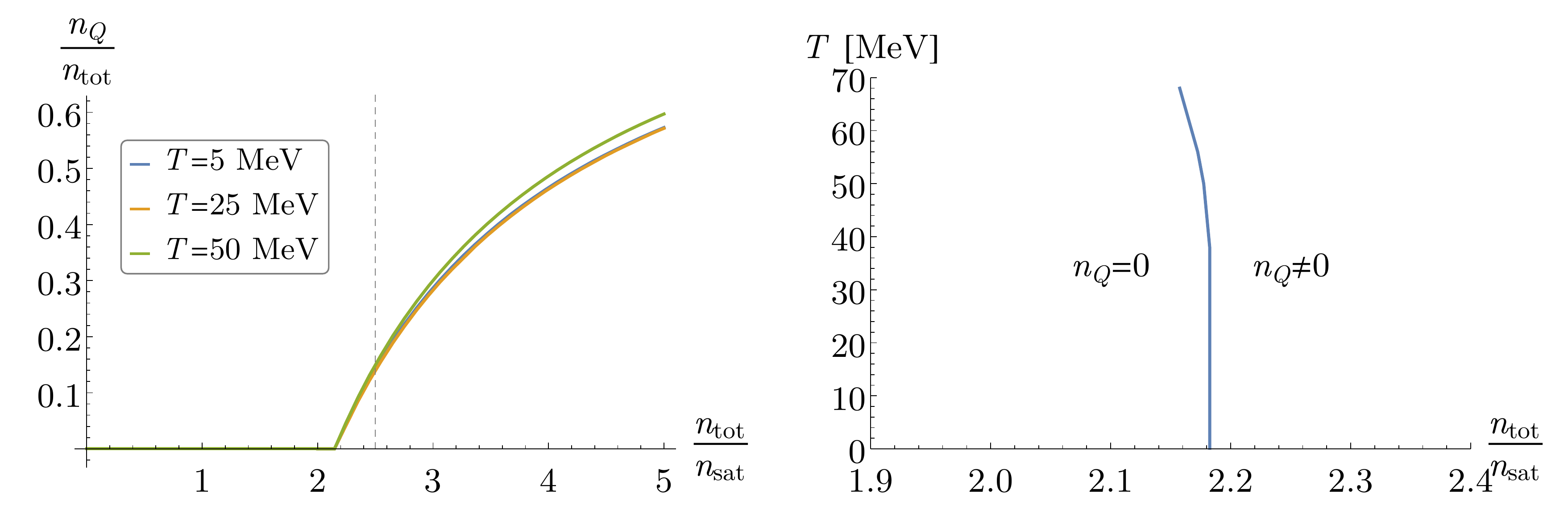}}
\caption{Left: Here we plot the quark fraction $n_Q/n_{tot}$ as a function of total density for a range temperatures found in neutron star mergers. The vertical dashed line indicates the hard core density. It can be seen that, as the temperature is increased, quarks are produced in greater amount. Right: Plotted is the quark content of the excluded volume model in the temperature density plane with a hard core density of $2.5 n_{sat}$. The onset density is largely insensitive to temperature decreasing slightly with increasing temperature under neutron star merger conditions.}
\label{fig:temp-onset-density}
\end{figure}

Next, in Fig. \ref{fig:temp-onset-density} we plot the quark fraction for a range of thermodynamic conditions. In particular, we choose temperatures and densities expected in neutron star mergers \cite{Perego:2019aa}, and once again the hard core density is taken to be $n_0=2.5 n_{sat}$. Several features of the excluded volume model can be gleaned from Fig. \ref{fig:temp-onset-density}. First, as expected from the low temperature expansion, quarks appear at lower density and in greater amount as the temperature is raised. Second, the onset density changes by only $\sim 2\%$ over the temperature range examined. This robustness of the onset density, while perhaps surprising, can be simply understood: quarks, residing within the shell of nucleons, are largely protected from the effects of temperature. Indeed, at the onset density the distribution function is nearly an all-nucleon configuration, and large temperatures are required to deform the Fermi surface enough to affect the quarks. For such a configuration, the factor $(1-n/n_0)^{-1/3}\sim N_c$ near the onset density, so the free energy shifts by an amount
\beq
\frac{F_N (T)-F_N(0)}{F_N(0)} = \frac{10}{9}\Big(\frac{2}{3\pi^2}\Big)^{1/3} \Big(\frac{M T}{n^{2/3}} \frac{1}{N_c^{2}}\Big)^2 ~
\label{eq:onset-estimate}
\eeq
at temperature is $T$. At $T=50$ MeV and $n=2.5 n_{sat}$, $\big(F_N (T)-F_N(0)\big)/F_N(0) \sim 3 \%$, which is close to the exact shift of $\sim2 \%$ seen in the onset density. Interestingly, the estimate \eq{eq:onset-estimate} reveals that $N_c$ factors are what cause the system be insensitive to temperature. Indeed, dimensional analysis alone would dictate that any thermal correction is set by the parameter $\sim \Big(M T /n_{tot}^{2/3}\Big)^2$, which is well beyond $\mathcal{O}(1)$ when $T=50$ MeV. 

In the right panel of Fig. \ref{fig:temp-onset-density}, we demarcate the quark content in various regions in the of the temperature/density plane. Within the model, we conclude that the effect of temperature is to slightly decrease the onset density relative to the zero temperature case. Whether this prediction of the excluded volume model extends to reality is not known, and must be checked with more detailed modeling of the nuclear micro-physics. Finally, varying the single parameter of this model, the hard core density, we find that the quark onset density remains well approximated by the simple relation $n_{onset} = 0.863 n_{0}$ obtained in \cite{Jeong:2019lhv}, correct to $5 \%$, for temperatures between $0 \text{ MeV}\leq T \leq 50$ MeV.

\section{Conclusion}
\label{conclusion}
In this paper we built a framework which can be used to compute thermal properties of quarkyonic matter within the excluded volume model. 
To do so, we introduced finite temperature distribution functions and entropy functionals motivated by the momentum-space shell structure of \cite{McLerran:2018hbz, Jeong:2019lhv}. Due to entropy, we find that the quark onset happens at lower densities at finite temperature than at zero temperature. Additionally, for a fixed density, we find that the amount of quarks produced increases as a function of temperature. If a nuclear to quark matter transformation behaves as suggested by this model, it would imply that relatively hot environments such as neutron stars mergers are more likely to carry signatures of quark matter than zero temperature environments. However, the thermal effects on the quark onset density is relatively small compared to what a naive dimensional estimate would suggest. It will be important to analyze the effect of temperature on other observables within this model to evaluate the full extent of thermal effects on quarkyonic matter.

It is also interesting to explore whether the excluded volume model for quarkyonic matter can be extended to include more realistic nuclear interactions at low density. A finite temperature analysis of such a model would be very useful in understanding whether thermal effects on the onset density depend on nuclear interactions within the scope of an excluded volume paradigm. We conclude by noting that, with finite temperature distribution functions and a complete set of thermodynamic functionals in hand, it is possible to compute in the excluded volume model a number of observables relevant to nuclear astrophysics. For example, the finite temperature equation of state as well as the speed of sound of quarkyonic matter can be computed. It may also be useful to compute mass-radius relations of neutron stars in the excluded volume model to compare with the constraints of \cite{Tews_2018, Capano:2019eae}, where a peak in the sound velocity between $2-3 n_{sat}$ is favored. Finally, using finite-temperature distribution functions extracted from the excluded volume model, neutrino scattering rates through hot, dense merger matter may be computed.

\section{Acknowledgements}
N.C.W. is grateful to Sanjay Reddy and Larry McLerran for their constructive comments. The work of N.C.W. was supported by U.S. DOE under Grant No. DE-FG02-00ER41132. The work of S.S. was supported by Iowa State University Startup funds.

\bibliography{quarkyon}
\end{document}